\documentclass[a4paper,draft]{amsart}
\usepackage{amsmath}
\usepackage{amssymb}
\usepackage{a4wide}
\usepackage{color}

\parskip\smallskipamount

\let\set\mathbb
\def\<#1>{\langle#1\rangle}

\def\e{\mathrm{e}}

\newtheorem{problem}{Problem}

\begin{document}

 \author[Shaoshi Chen]{Shaoshi Chen\,$^\ast$}
 \address{Shaoshi Chen, KLMM, AMSS, Chinese Academy of Sciences, China.}
 \email{schen@amss.ac.cn}
 \thanks{$^\ast$ Supported by the NSFC grant 11501552 and
   by the President Fund of the Academy of Mathematics and Systems Science, CAS (2014-cjrwlzx-chshsh).
   This work was also supported by the Fields Institute's 2015 Thematic Program on Computer Algebra.}

 \author[Manuel Kauers]{Manuel Kauers\,$^\ast$}
 \address{Manuel Kauers, Institute for Algebra, J. Kepler University Linz, Austria.}
 \email{manuel.kauers@jku.at}
 \thanks{$^\ast$ Supported by the Austrian FWF grants F5004, Y464-N18, and W1214.}

 \title{Some Open Problems related to Creative Telescoping}

 \begin{abstract}
   Creative telescoping is the method of choice for obtaining information about
   definite sums or integrals. It has been intensively studied since the early
   1990s, and can now be considered as a classical technique in computer algebra.
   At the same time, it is still subject of ongoing research. In this paper, we
   present a selection of open problems in this context. We would be curious
   to hear about any substantial progress on any of these problems.
 \end{abstract}

 \maketitle

 \section{Introduction}

 Summation problems arise in all areas of mathematics, especially in discrete mathematics
 and combinatorics. The general task is to compute for a given expression describing
 a summand sequence $f(n,k)$ an expression that describes the sum sequence $F(n)=\sum_{k=0}^n f(n,k)$.
 Depending on the type of expressions allowed for summand and/or sum, a solution may or may
 not exist. The classical class of expressions considered in the theory of symbolic summation
 is the class of hypergeometric terms. A univariate sequence $f(k)$ is called hypergeometric if the
 shift quotient $f(k+1)/f(k)$ can be simplified to a rational function in~$k$. For example,
 $f(k)=k!$ is hypergeometric because $f(k+1)/f(k)=k+1$ is a polynomial.
 Another example is $f(k)=2^k$.
 Gosper's algorithm~\cite{gosper78} solves the decision problem for hypergeometric summation:
 given a hypergeometric term $f(k)$ (i.e., given a rational function $r(k)$ such that
 $f(k+1)/f(k)=r(k)$), it computes a hypergeometric term $F(k)$ such that $F(k+1)-F(k)=f(k+1)$,
 or it certifies that no such hypergeometric term exists.
 When $F(k)$ is found, it implies the closed form representation $\sum_{k=0}^n f(k)=F(n+1)-F(0)$.
 For example, Gosper's algorithm can find the formula $\sum_{k=0}^n k\,k!=(n+1)!-1$.

 Gosper's algorithm only applies to so-called indefinite sums. These are sums in which the upper
 summation bound is a variable that does not occur in the summand expression. All other sums
 are called definite. For example,
 $\sum_{k=0}^n \binom mk$ is an indefinite sum (involving a parameter~$m$), while
 $\sum_{k=0}^n \binom nk$ is a definite sum.
 The distinction is important because there does exist a closed form for the latter sum
 (it is equal to the nice expression~$2^n$), but no closed form exist when $m$ and $n$ are
 unrelated.

 In order to process definite sums, we can use the technique of creative telescoping.
 Informally, creative telescoping solves the following problem:
 Given an expression $f(n,k)$, it computes polynomials $c_0(n),\dots,c_r(n)$, not all zero,
 and an expression~$g(n,k)$, such that
 \[
   c_0(n)f(n,k) + \cdots + c_r(n)f(n+r,k) = g(n,k+1) - g(n,k).
 \]
 When such a relation is available, we can sum it for $k$ from $0$ to $n$ to obtain a relation of the form
 \[
   c_0(n) F(n) + \cdots + c_r(n) F(n+r) = G(n)
 \]
 for the definite sum $F(n)=\sum_{k=0}^n f(n,k)$ and some explicit expression~$G(n)$.
 From such an equation, other algorithms can be used to find closed form representations for $F(n)$ (or
 prove that there are none), or information about its asymptotic behaviour for $n\to\infty$, or
 to compute a large number of terms of the sequence efficiently.

 The method of creative telescoping was propagated by Zeilberger in the early
 1990s~\cite{wilf90,wilf92a,zeilberger91,petkovsek97}  (although the word ``creative telescoping'' already appears
 in~\cite{vanDerPoorten79}).  Zeilberger also gave the first algorithm for creative
 telescoping applicable to hypergeometric terms. This algorithm, now known as
 Zeilberger's algorithm, is a clever modification of Gosper's algorithm. Zeilberger
 also formulated a vision for doing creative telescoping in the much more
 general realm of holonomic functions~\cite{zeilberger90}. Over the years, this led to the
 development of operator-based techniques such as Chyzak's algorithm~\cite{chyzak98a,chyzak00}
 as well as difference-field-based techniques mainly developed by
 Schneider~\cite{schneider01,schneider08,schneider13b}.

 \def\ann{\operatorname{ann}}%
 Ore algebras provide a setting in which the creative telescoping problem can be formulated
 in great generality. To give an idea, let us consider the case where $C$ is a field of
 characteristic zero,
 $K=C(n,k)$ is the field of rational functions in $n$ and $k$ with coefficients in~$C$, and
 $A=K[S_n,S_k]$ is the polynomial ring in two variables $S_n$,~$S_k$ with coefficients in~$K$.
 The multiplication on $A$ is defined in such a way that we have
 $S_nS_k = S_kS_n$ and
 $S_n r(n,k)=r(n+1,k)S_n$ and
 $S_k r(n,k)=r(n,k+1)S_k$ for all $r\in K$.
 The elements of $A$ can then be viewed as operators that act on a space $F$ of bivariate sequences.
 For any particular sequence $f\in F$, we may then consider the left ideal $\ann(f)=\{L\in A:L\cdot f=0\}$
 of all the operators in $A$ which map $f$ to zero. Then the problem of creative telescoping is
 to find some operator $P\in C(n)[S_n]\setminus\{0\}$ and some operator $Q\in A$ such that $P-(S_k-1)Q\in\ann(f)$.
 In such a representation, $P$~is called a telescoper for $f$ and $Q$ is called a certificate for~$P$.

 There are some other flavors of the creative telescoping problem which are also of interest.
 In particular, there is a differential version, which is useful for integration. In this case,
 we consider the Ore algebra $A=C(x,y)[D_x,D_y]$ consisting of all linear differential operators
 with coefficients in $C(x,y)$, acting on a space $F$ of bivariate functions.
 Note that the multiplication laws for differential operators are slightly different from the multiplication
 laws for recurrence operators: here we have $D_xD_y=D_yD_x$ and $D_xa=aD_x + \frac d{dx}a$ and $D_ya=aD_y+\frac d{dy}a$
 for all $a\in C(x,y)$.
 For any particular function $f\in F$, let $\ann(f)=\{L\in A:L\cdot f=0\}$ again denote the left ideal
 consisting of all the operators in $A$ that map $f$ to zero. The problem of creative telescoping is then
 to find some operator $P\in C(x)[D_x]\setminus\{0\}$ and some operator $Q\in A$ such that $P-D_yQ\in\ann(f)$.
 In the context of integration, such an operator can serve the same purpose as a creative telescoping relation
 of the form discussed before in the context of summation:
 From $(P-D_yQ)\cdot f=0$ follows $0=\int_0^1((P-D_yQ)\cdot f)(x,y)dy=P\cdot\int_0^1 f(x,y)dy-[(Q\cdot f)(x,y)]_{y=0}^{1}$,
 so we have $P\cdot F(x)=G(x)$ for $F(x)=\int_0^1 f(x,y)dy$ and some simple and explicit function~$G(x)$.

 A lot of research has been done on algorithms for creative telescoping during
 the past 25~years. A reasonably complete and almost up-to-date overview of the
 state of the art is given in Chyzak's Habilitation thesis from
 2014~\cite{chyzak14}. The focus of this thesis is on the algorithmic aspects
 and the theoretical foundations. In addition, there are many papers that
 implicitly or explicitly make use of the theory by simplifying sums or
 integrals using computer programs based on the method of creative
 telescoping. This underlines the importance of the method. At the same time,
 despite the successful work on creative telescoping that has been done in the
 past, there is still a number of open problems which do not yet have
 satisfactory answers. In the present article, we offer a collection of such
 open problems. The choice is obviously biased by our personal interests.
 However, we believe that significant progress on any of these problems would be
 a valuable contribution to the advance of symbolic summation.

\section{Reduction-Based Algorithms}

 Algorithms for creative telescoping can be distinguished according to their input
 class or according to the algorithmic technique they are based on. The available
 algorithmic techniques can be divided into four generations of creative telescoping algorithms.
 Algorithms from the first generation use elimination theory for operator ideals~\cite{fasenmyer49,takayama90,takayama90a,petkovsek97,wegschaider97,chyzak98}.
 Zeilberger's algorithm from 1990~\cite{zeilberger90a} and its generalizations~\cite{almkvist90,chyzak00,kauers07n,schneider13b} form the second generation.
 The third generation is based on an idea that was first formulated by Apagodu and Zeilberger~\cite{mohammed05,apagodu06}
 and has later been refined and generalized~\cite{koutschan10b,chen12b,chen12c,chen14a}.
 Algorithms from the fourth and most recent generation of creative telescoping algorithms are called
 reduction-based algorithms. They were first introduced by Bostan et al.~\cite{bostan10b} for
 integration of rational functions. The basic idea is as follows. Consider a rational
 function $f=p/q\in C(x,y)$. The task is to find $c_0,\dots,c_r\in C(x)$ such that there
 exists $g\in C(x,y)$ with
 \[
   c_0 f + c_1 D_x f + \cdots + c_r D_x^r f = D_y g.
 \]
 Consider the partial derivatives $f,D_xf,D_x^2f,\cdots\in C(x,y)$. Using Hermite reduction,
 we can write each of them in the form $D_x^i f = D_y g_i + h_i$ for some $g_i,h_i\in C(x,y)$
 where $h_i$ has a square free denominator whose degree exceeds the degree of its numerator.
 The denominators of all these $h_i$ divide the square free part of the denominator of~$f$
 in $C(x)[y]$, so the $C(x)$-subspace of $C(x,y)$ generated by $h_0,h_1,\dots$ has finite
 dimension. If the dimension is~$r$, then we can find $c_0,\dots,c_r\in C(x)$, not all zero,
 such that $c_0h_0+\cdots+c_rh_r=0$. For these $c_0,\dots,c_r$ we then have
 \[
   c_0 f + c_1 D_x f + \cdots + c_r D_x^r = D_y (g_0 + g_1 + \cdots + g_r) + 0,
 \]
 as desired.

 The approach is not limited to rational functions and has been generalized to
 hyperexponential terms~\cite{bostan13a}, hypergeometric terms (for the summation
 case)~\cite{chen15a,huang16} and algebraic functions~\cite{chen16a}. It has also been worked
 out for the mixed case when the integrand is a hyper\-geo\-metric-hyper\-expo\-nen\-tial
 term $f_n(x)$~\cite{bostan16},
 and it is being worked out by Du, Huang and Li~\cite{li16}
 the $q$-case. At this stage, the summation case for hypergeometric-hyperexponential
 terms $f_n(x)$ is still open, so this shall be our first problem.

 \begin{problem}
   Develop a reduction based creative telescoping algorithm which for a given hyper\-geo\-me\-tric-hyperexponential
   term $f_n(x)$ computes, if possible, rational functions $c_0,\dots,c_r\in C(x)$, not all zero,
   such that there exists a hypergeometric-hyperexponential term~$g_n(x)$ with
   \[
    c_0 f_n(x) + \cdots + c_r D_x^r f_n(x) = g_{n+1}(x) - g_n(x).
   \]
 \end{problem}

 In the pure differential case, we could consider integrands from larger classes of
 functions. The largest class considered so far was the class of algebraic functions~\cite{chen16a}.
 It is based on Trager's Hermite reduction~\cite{trager84,bronstein98}. The correctness of the method relies
 heavily on Chevalley's theorem~\cite{chevalley51}, according to which any non-constant algebraic function must
 have a pole at some place (possibly over infinity). Since there is no analogous theorem
 for general D-finite functions, not even for solutions of Fuchsian equations, it is not clear
 how to generalize the reduction based algorithm from algebraic functions to (Fuchsian) D-finite
 functions. This is our second problem.

 \begin{problem}
   Develop a reduction based creative telescoping algorithm which for a given (Fuchsian) D-finite
   function $f(x,y)$ computes, if possible, rational functions $c_0,\dots,c_r\in C(x)$, not all zero, such
   that there exists an operator $Q\in C(x,y)[D_x,D_y]$ with $(c_0+c_1D_x+\cdots+c_rD_x^r)\cdot f = D_y Q\cdot f$.
 \end{problem}

 \section{Order-Degree Curves}

 When a function admits a telescoper, the telescoper is not uniquely determined. The set of telescopers
 rather forms a left ideal in the operator algebra $C(x)[D_x]$ (or in $C(n)[S_n]$, respectively).
 Since the operator algebras $C(x)[D_x]$ and $C(n)[S_n]$ are left-Euclidean domains, it follows that
 there is a unique monic telescoper of minimal possible order---called the minimal telescoper---and that
 all the other telescopers are left-multiples of this telescoper.

 For the purpose of estimating the computational cost of creative telescoping algorithms, it is interesting
 to know bounds for the size of telescopers relative to characteristic parameters of the input.
 Besides bounds on the order~$r$ of the telescopers, it is also of interest to bound the sizes of its
 coefficients. After clearing denominators (from left), we can assume that the telescoper lives in
 $C[x][D_x]$ or~$C[n][S_n]$, and we can ask for its degree~$d$ with respect to $x$ or~$n$.

 Unlike $C(x)[D_x]$ and $C(n)[S_n]$, the rings $C[x][D_x]$ and $C[n][S_n]$ are not left-principal. As a
 consequence, we can in general not minimize the order $r$ and the degree $d$ simultaneously. Instead,
 we must expect that telescopers of low order $r$ have a high degree~$d$ and telescopers of low degree~$d$
 have high order~$r$. To describe the general situation, we use a function $c\colon\set R\to\set R$ such
 that for each $r\geq r_{\min}\in\set N$ there is a telescoper of order~$r$ and degree at most~$c(r)$.
 The graph of the function~$c$ is called an order-degree curve for the summation/integration problem at hand.

 It turns out that order-degree curves can be derived from the Apagodu-Zeilberger algorithm~\cite{mohammed05}.
 Apagodu and Zeilberger used their approach to derive bounds on the order of the telescopers. Again, the idea
 is easily explained for the case of rational functions. Consider $f=p/q\in C(x,y)$ and suppose for simplicity
 that $\deg_y p<\deg_y q$. By induction, it can be shown that $D_x^i f = p_i/q^{i+1}$ for some polynomial
 $p_i\in C(x)[y]$ of degree at most $\deg_y p_i\leq \deg_y p + i\deg_y q$. Therefore, for any choice $r\in\set N$
 and any choice $c_0,\dots,c_r\in C(x)$, we have that $c_0 f + c_1 D_x f + \cdots + c_r D_x^r f$ is a rational
 function with denominator~$q^{r+1}$ and a numerator whose degree is bounded by $\deg_y p+r\deg_y q$.
 Now consider a rational function $g=u/q^r$ with $u=u_0+u_1y+\cdots+u_sy^s\in C(x)[y]$.
 Then $D_y g=v/q^{r+1}$ for some $v\in C(x)[y]$ of degree at most $s + \deg_yq$.
 In order to get the desired equality $c_0 f + c_1 D_x f + \cdots + c_r D_x^r f=D_y g$, we multiply both
 sides by $q^{r+1}$ and equate coefficients with respect to~$y$.
 This gives a linear system over $C(x)$ for the variables $c_0,\dots,c_r,u_0,\dots,u_s$.
 These are $(r+1)+(s+1)$ variables.
 The number of equations is at most $1+\max(\deg_y p+r\deg_y q,s+\deg_q)$, which simplifies
 to $1+\deg_y p+r\deg_y q$ if we choose $s=\deg_y p+(r-1)\deg_y q$.
 The number of variables exceeds the number of equations if $(r+1)+(\deg_y p+(r-1)\deg_y q+1)>1+\deg_y p+r\deg_y q$,
 i.e., if $r>\deg(q)-1$.
 It follows that for $r\geq\deg(q)$ the linear system will have a nontrivial solution.
 For this nontrivial solution, at least one of $c_0,\dots,c_r,u_0,\dots,u_s$ is nonzero.
 It is then not possible that $c_0,\dots,c_r$ are all zero, because by our simplifying assumption $g$ is a rational
 function whose numerator as lower degree than its denominator, so $D_yg$ can only be zero if $g$ is zero, and then
 also $u_0,\dots,u_s$ would all have to be zero.
 We have thus shown that the minimal order telescoper for $f$ has order at most~$\deg(q)$.

 The reasoning can be refined such as to also provide bounds for the degrees of the telescopers. This has
 been done for hyperexponential terms in~\cite{chen12b} and for hypergeometric terms in~\cite{chen12c}. The resulting curves
 are simple hyperbolas. However, the degree bounds are not sharp. For the hypergeometric case, also the
 bit size of the integer coefficients has been analyzed~\cite{kauers14d}.
 For general D-finite functions, we know bounds for the order of the telescopers but an order-degree curve
 has not yet been worked out. Therefore:

 \begin{problem}
   Derive an order-degree-curve for general D-finite functions.
 \end{problem}

 It would also be interesting to have bounds for the bit size not only for hypergeometric input but also for
 other classes, for example for hyperexponential terms.

 \begin{problem}
   Derive bounds for the bit size of telescopers for hyperexponential terms.
 \end{problem}

 Experiments show that the order-degree curves following from the analysis of Apagodu-Zeilberger-like algorithms
 are not sharp. Better bounds could be obtained if we had a better understanding of the singularities of
 telescopers. It was shown in~\cite{jaroschek13a} how the distinction between removable and non-removable singularities of
 an operator $L\in C[x][D_x]$ implies a curve that very accurately describes the degrees of the elements of
 $(C(x)[D_x]L)\cap C[x][D_x]$. Here, a singularity of $L$ is defined as a root of the leading coefficient
 polynomial (the coefficient of the highest derivative), and such a singularity is called removable if there
 exists an operator $Q\in C(x)[D_x]$ such that $QL$ is in $C[x][D_x]$ and does not have this singularity.
 The terminology is analogous for recurrence operators, and the connection to order degree curves observed
 in \cite{jaroschek13a} also applies to this case.

 Several algorithms are known for identifying the removable singularities of an operator~\cite{abramov99b,abramov06a,chen15b}.
 Therefore, when a telescoper is known, we obtain a very accurate order-degree curve.
 However, for the design of efficient creative telescoping algorithms it would be useful to have order-degree
 curves that can be easily read off from the summand/integrand, rather than from the telescoper.
 The question therefore is whether it is possible to predict the removable and non-removable singularities
 of a telescoper directly from the summand/integrand. This leads to the next problem.

 \begin{problem} \textbf{\emph{(a)}}
   Find a way to determine the removable and non-removable singularities of a telescoper for a given proper
   hypergeometric term $f(n,k)=p c^n d^k \prod_{i=1}^m \Gamma(\alpha_i n+\beta_i k+\gamma_i)^{e_i}$
   ($p\in C[n,k]$, $c,d\in C$, $\alpha_i,\beta_i,\e_i\in\set N$, $\gamma_i\in C$),
   using less computation time than needed for computing a telescoper.

   \textbf{\emph{(b)}} The analogous question for hyperexponential terms $f(x,y)=\exp(a/b)\prod_{i=1}^m p_i^{c_i}$
   ($a,b,p_i\in C[x,y],c_i\in C$).
 \end{problem}

 \section{Differential and Difference Fields}\label{sec:2}

 In the area of differential algebra, a pair $(K,d)$ is called a differential
 field if $K$ is a field and $d\colon K\to K$ is such that $d(a+b)=d(a)+d(b)$
 and $d(ab)=d(a)b+ad(b)$ for all $a,b\in K$. For example, the field $K=C(x)$ of
 rational functions forms a differential field together with the usual
 derivation~$\frac d{dx}$.  More generally, appropriate differential fields can
 be used to emulate the behaviour of expressions involving elementary functions
 under differentiation. The corresponding differential fields are called
 liouvillean fields. They are used in Risch's integration
 algorithm~\cite{risch69,risch70,bronstein97,bronstein98}. Analogously, a difference field is a pair $(K,s)$
 where $K$ is a field and $s\colon K\to K$ is such that $s(a+b)=s(a)+s(b)$ and
 $s(ab)=s(a)s(b)$ for all $a,b\in K$, i.e., $s$ is an automorphism. Difference
 fields corresponding to liouvillean fields are called $\Pi\Sigma$-fields.
 They emulate the behaviour of expressions involving nested
 sums and products under shift and are used in Karr's summation
 algorithm~\cite{karr81,karr85}.

 The creative telescoping problem can be formulated for differential and
 difference fields.  In the differential case, let $K$ be a field with two
 derivations $d_x,d_y\colon K\to K$ that commute with each other, and consider
 the operator algebra $A=K[\partial_x,\partial_y]$ with the commutation rules
 $\partial_x\partial_y=\partial_y\partial_x$ and $\partial_x a = a\partial_x +
 d_x(a)$ and $\partial_y a = a\partial_y + d_y(a)$ for all $a\in K$. Such an
 operator algebra may act on some function space~$F$. For a given $f\in F$ we
 may then ask, like before, whether there exists $P,Q\in A$ such that
 $(P-\partial_yQ)\cdot f=0$. Here, $P$~must belong to $K_x[\partial_x]$, where
 $K_x=\{u\in K:d_y(u)=0\}$ is the subfield of $K$ consisting of all elements of
 $K$ that are constant with respect to~$y$. The version for difference fields
 is analogous.

 Schneider~\cite{schneider01} was the first to observe that Karr's summation algorithm
 can be used to solve the creative telescoping problem in very much the same way
 as Gosper's algorithm is exploited in Zeilberger's algorithm. He has been working
 on refinements, extensions, and generalizations of summation technology based on
 difference field theory for many years and has obtained spectacular results, see
 \cite{schneider13b} and the references given there. Yet, some questions have not yet
 been addressed. In particular, there is no general theory which clarifies
 under which circumstances a telescoper exists (a question that is settled for the
 classical hypergeometric case by the work of Abramov et al.~\cite{abramov02a,abramov02,abramov03,abramov05a}), or to
 give a priori bounds on their order or on the cost for their computation. Similar
 remarks apply in the differential case, for which Raab~\cite{raab12} has recently
 formulated a creative telescoping approach based on Risch's algorithm, but no
 theoretical results concerning existence or size of telescopers were given.

 \begin{problem}
   For the creative telescoping problem over liouvillean fields (in the differential case)
   or for $\Pi\Sigma$-fields (in the shift case), derive a criterion for the existence
   of a telescoper. For the cases where telescopers exist, derive bounds on their order.
 \end{problem}

 In contrast to D-finite functions in the differential case, elementary
 functions may not have a telescoper. One obstruction to the existence of a telescoper
 may be the fact that an elementary function can only be elementary integrable if
 all its residues are constant (cf. Section~5.6 of~\cite{bronstein97}). A telescoper
 must therefore at least map all the residues of the given function to constants. This
 is only possible if the residues are D-finite, which may not be the case. For example,
 the function $f(x,y)=\frac{x}{(\e^x-1)(1-y)}$ cannot have a telescoper with respect to~$y$,
 because its residue at $y=1$ is $\frac{x}{\e^x-1}$, which is not D-finite.

 For the shift case, Schneider has an algorithm~\cite{schneider05a} which computes for a given
 nested sum expression an equivalent expression in which the nesting depth is as small
 as possible. This is remarkable because the equivalent representation with minimal
 depth does usually not belong to the same field in which the input sum is given. So far
 there is no analogous algorithm for the differential case, although it would be
 interesting to have one. Therefore:

 \begin{problem}
   Design an algorithm which finds for a given expression of nested indefinite integrals
   an equivalent expression for which the the nesting depth is as small as possible.
 \end{problem}


 Our last problem in this section relates to the structure of the class of elementary functions.
 As this class is not closed under integration, the set of elementary integrable elementary functions
 forms a proper subclass. This class in turn contains integrable as well as non-integrable functions.
 It is clear that for every $n\in\set N$, there is an elementary function which is $n$ times elementary
 integrable but not $n+1$ times. An example is the $n$th derivative of~$\e^{x^2}$. On the other hand, there
 are also elementary functions which can be integrated arbitrarily often without ever leaving the class of
 elementary functions, for example polynomials. What other functions have this property?

 \begin{problem}
   Determine the class of elementary functions with the property that for every
   $n\in\set N$, their $n$-fold integral is again elementary.
 \end{problem}

 Using repeated partial integration, we can show that a function $f(x)$ belongs to this class if and only
 if for every $n\in\set N$ the function $x^n f(x)$ is elementary integrable. This implies that all rational
 functions are arbitrarily often elementary integrable. Note that this is not obvious because the integral
 of a rational function may involve logarithms of algebraic functions, and such functions need not be
 elementary integrable.

 \section{The Multivariate Case}

 While most single sums appearing in practical applications are nowadays no
 challenge for a computer algebra system, multiple sums may still be too
 hard. One natural reason is that multiple sums tend to involve expressions in
 many variables, and such expressions can quickly become too large to be handled
 efficiently. Another reason is that the algorithms we know for single sums are
 better than those we know for multiple sums. For single sums, Zeilberger's
 algorithm supersedes elimination methods such as the so-called Sister Celine
 algorithm~\cite{fasenmyer49,verbaeten74,petkovsek97}. But while the algorithm
 of Sister Celine has been generalized to multisums~\cite{wilf92a,wegschaider97}, there
 is no multivariate Zeilberger algorithm yet. We do not even know a multivariate
 Gosper algorithm.

 \begin{problem}
   Develop an algorithm which takes as input a multivariate hypergeometric term
   $h$ in $m$ discrete variables $k_1,\dots,k_m$, and decides whether there exist
   hypergeometric terms $g_1,\dots,g_m$ such that
   \[
     h = \Delta_1(g_1) + \cdots + \Delta_m(g_m).
   \]
   Here, $\Delta_i$ is the forward difference operator with respect to
   the variable~$k_i$, i.e., $\Delta_i
   f(k_1,\dots,k_m)=f(k_1,\dots,k_i+1,\dots,k_m)-f(k_1,\dots,k_i,\dots,k_m)$.
 \end{problem}

 A solution of this problem would be an important step towards the development
 of a Zeilberger-like algorithm for multisums. Recently, Chen and
 Singer~\cite{chen12d,chen14b} have given a necessary and sufficient condition
 for the case when $h$ is a rational function in two variables. Their criterion
 was then turned into an algorithm by Hou and Wang~\cite{hou15}.
 In~\cite{chen16} these results were used to derive some conditions on the
 existence of telescopers for trivariate rational functions. Summability
 criteria for larger classes, such as the class of hypergeometric terms, may
 analogously allow for the formulation of existence criteria for telescopers in
 the multivariate setting. In the long run, we would hope that a multivariate Gosper algorithm
 serves as a starting point for the development of a reduction-based
 creative telescoping algorithm for the multivariate setting.

 The corresponding question for bivariate rational functions in the differential
 case has been studied already by Picard~\cite{picard06,picard33} many years
 ago. More recently, Griffiths and Dwork~\cite{dwork62,dwork64,griffiths69,griffiths69a} gave a method
 that works for any number of variables but requires some kind of regularity of the denominator. An
 algorithm for creative telescoping based on these results was given by Bostan
 et al.~\cite{bostan13b}.

 \section{Binomial Sums}

 The principal application of creative telescoping is the construction of
 recurrence relations satisfied by definite sums. As already indicated in the introduction,
 such a recurrence can be obtained from a telescoper-certificate pair for the summand.
 However, some care is necessary for this step. In order to be able to sum a relation
 \[
   c_0(n)f(n,k)+\cdots+c_r(n)f(n+r,k) = g(n,k+1)-g(n,k)
 \]
 for $k$ from $0$ to~$n$, we must assure that the right hand side involving the certificate
 $g(n,k)$ does not have any poles for the values $k$ in this range. Unfortunately, such
 poles do appear in examples, and although they usually cancel each other nicely, it is not
 easy to verify this algorithmically. See \cite{chyzak14a} for a detailed case study in
 this context.

 For indefinite hypergeometric single sums, Abramov and Petkovsek~\cite{abramov05}
 discuss an alternative to Gosper's algorithm that handles special points
 properly. Ryabenko~\cite{ryabenko11} gives an accurate summation algorithm for
 definite sums over a particular class of hypergeometric terms. A continuation
 of her work towards the full class of hypergeometric terms (or even beyond)
 would be worthwhile.

 \begin{problem}
   Develop an algorithm that correctly transforms a telescoper-certificate pair
   for a hypergeometric term into a recurrence for the corresponding definite
   sum.  In particular, the algorithm should property take care of any possible
   issues arising from poles in the certificate.
 \end{problem}

 It appears that the situation is somewhat easier for summands with compact
 support. A hypergeometric term $f(n,k)$ is said to have compact support if for
 every $n\in\set Z$ there are only finitely many $k\in\set Z$ such that $f(n,k)$
 is different from zero. In this case, the infinite sum $\sum_{k=-\infty}^\infty
 f(n,k)$ is in fact a terminating sum. For example, we have $\sum_{k=0}^n\binom
 nk=\sum_{k=-\infty}^\infty\binom nk$ because $\binom nk=0$ when $k<0$ or $k>n$.

 When the sum over $k$ runs through all integers (and there are no issues with
 poles in the certificate), the transformation of a telescoper-certificate pair
 to a recurrence for the definite sum is particularly nice. One reason is that
 the operator $\sum_{k=-\infty}^\infty$ commutes with the shift operator~$S_n$,
 and therefore, with the telescoper. A second reason is that the right hand side
 $\sum_{k=-\infty}^\infty\bigl(g(n,k+1)-g(n,k)\bigr)$ invariably collapses to
 zero (because when $f(n,k)$ has compact support, then so does
 $g(n,k)$). Therefore, in the case of compact support, the telescoper for
 $f(n,k)$ is precisely the recurrence for $\sum_k f(n,k)$.

 Viewing hypergeometric terms as algebraic objects, it is somewhat
 unsatisfactory that the concept of compact support is defined ``analytically''
 in terms of the values of sequences associated to the terms. In view of a
 possible automation, a more algebraic explanation of the phenomenon would be
 useful. A finite summation operator such as $\Sigma:=\sum_{k=0}^n$ does not
 commute with the shift~$S_n$. However, if we introduce the evaluation
 operator~$E_n$ that acts on bivariate terms by setting $k$ to~$n$, then we have
 the commutation rule $\Sigma S_n = S_n\Sigma - E_nS_n$. This rule expresses
 the fact $\sum_{k=0}^n f(n+1,k)=\sum_{k=0}^{n+1}f(n+1,k) - f(n+1,n+1)$.
 Now consider a telescoper $P=c_0(n)+c_1(n)S_n+\cdots+c_r(n)S_n^r$ with a
 corresponding certificate~$Q$, so that $(P - \Delta_k Q)\cdot f(n,k)=0$. Applying $\Sigma$ to
 this relation and using the commutation rules leads to
 \[
 P\Sigma\cdot f(n,k) =
 \biggl(\biggl(c_1(n)E_nS_n + \cdots + c_r(n)\sum_{i=1}^r S_n^{r-i}E_nS_n^i\biggr) + (E_nS_nQ - E_0Q)\biggr)\cdot f(n,k),
 \]
 where $E_0$ denotes an evaluation operator that sets $k$ to~$0$.
 We see that the telescoper $P$ translates directly into an annihilating operator for the
 sum if and only if the right hand side is zero, i.e., if the operator on the right annihilates the summand.
 Note that it is irrelevant whether $f(n,k)$ has compact support.

 For the differential case, Regensburger, Rosenkranz and collaborators have
 developed a theory of operator algebras that include both derivations as well
 as integration operators. Their principal motivation is to solve boundary value
 problems,
 see~\cite{rosenkranz08,rosenkranz08a,regensburger09,guo14,regensburger16} and
 the references given there for an overview of their results. Their algebras
 also contain evaluation operators similar to the $E_n$ introduced above. We
 would like to see an analogous theory for operator algebras involving summation
 as well as shift operators.

 \begin{problem}
   Develop a theory of operator algebras including shift as well as summation operators,
   analogous to the theory of Regensburger and Rosenkranz.
   In this theory, find an algebraic explanation why the right hand side of a creative
   telescoping relation often vanishes for binomial sums.
 \end{problem}

 In a recent paper, Bostan et al.~\cite{bostan15} approach the problems related to boundary
 conditions and possible poles in the certificate from a different direction. Instead of applying
 creative telescoping directly to the sum in question, they translate the summation
 problem into an integration problem and apply creative telescoping to this problem.
 One advantage of this approach is that for the resulting contour integrals
 there are no problems related to singularities, because the path of integration can
 always be deformed such as to avoid all the singularities. For this reason, it is not
 necessary to inspect the certificate, and it is possible to employ efficient algorithms
 which only compute the telescoper.
 So far the approach does not apply to all hypergeometric sums but only to a subclass. They
 call it the class of \emph{binomial sums} and they show for the case of one variable
 that a sequence is a binomial sum (in the sense of their definition) if and only if it
 is the diagonal of a multivariate rational function. The diagonal of a multivariate
 power series $\sum_{n_1,\dots,n_r=0}^\infty a_{n_1,\dots,n_r}x_1^{n_1}\cdots x_r^{n_r}$
 is defined as the univariate series $\sum_{n=0}^\infty a_{n,n,\dots,n}x^n$.
 The definition of binomial sums also covers sums with several variables, but no
 characterization of binomial sums in several variables is given in~\cite{bostan15}.

 \begin{problem}
   Prove or disprove:
   A multivariate sequence $(b_{k_1,\dots,k_s})$ in $s$ discrete variables is a binomial
   sum in the sense of~\cite{bostan15}
   if and only if there exists a rational power series
   \[\sum_{n_1,\dots,n_r=0}^\infty a_{n_1,\dots,n_r}x_1^{n_1}\cdots x_r^{n_r}\]
   and $i_1,\dots,i_s\in\set N$ with $i_1+\cdots+i_s=r$ such that
   for all $k_1,\dots,k_s$ we have
   \[
   b_{k_1,\dots,k_s}=a_{
     \underbrace{\scriptstyle k_1,\dots,k_1}_{i_1},
     \underbrace{\scriptstyle k_2,\dots,k_2}_{i_2},\dots,
     \underbrace{\scriptstyle k_s,\dots,k_s}_{i_s}}
   \]
 \end{problem}

 An important open problem in the context of diagonals is Christol's
 conjecture~\cite{christol88}, which says that every formal power series with integer
 coefficients and a positive radius of convergence which is the solution of a
 linear differential equation with polynomial coefficients is the diagonal of
 some rational power series. In this conjecture, no statement is made about the
 number of variables of the rational power series. Bostan et al.~\cite{bostan15}
 remark that we must at least allow for three variables, and that no explicit
 example is known which requires more.

 Because of its connection to diagonals, the class of binomial sums as introduced in~\cite{bostan15}
 is not as artificial as it seems at first glance. Nevertheless, also a natural restriction
 is a restriction. It would be interesting to extend the applicability of the algorithm to
 a wider class.

 \begin{problem}
   Generalize the algorithm of~\cite{bostan15} from binomial sums to arbitrary hypergeometric
   sums.
 \end{problem}

%

 \section{Nonlinear Equations and Annihilators of Positive Dimension}

 In the theory of ``holonomic systems''~\cite{zeilberger90}, summands and
 integrands are represented by ideals of operators by which they are
 annihilated. Properties of the ideal are used to ensure the existence of
 telescopers and the termination of algorithms. A condition that is typically
 imposed is that the ideal has Hilbert dimension~$0$. In this case, the
 annihilated function is called D-finite.  Many functions of practical relevance
 happen to be D-finite, but it is natural to ask to whether D-finiteness is
 really needed for creative telescoping to succeed. It turns out that it is not.
 Already in the 1990s, Majewicz has given a variant of creative telescoping
 applicable to Abel-type identities~\cite{majewicz96}. The key observation is
 that such identities exist because the sum has more than one free variable, and
 this can compensate for the lack of relations preventing the summand from being
 D-finite. A summation algorithm by Kauers~\cite{kauers07n} for sums involving
 Stirling numbers and an algorithm by Chen and Sun~\cite{chen09} for sums
 involving Bernoulli numbers are based on similar observations. In 2009, the
 phenomenon was formulated in more general terms by Chyzak et
 al.~\cite{chyzak09a}. They showed that telescopers can exist also when the
 annihilator of the summand/integrand has positive dimension. More precisely,
 consider a function with $n$ free variables and $k$ summation/integration
 variables, let $I$ be the annihilator of the function and let $T$ be the ideal
 of telescopers (in the smaller operator algebra corresponding only to the $n$
 free variables). Then they show that $\dim T\leq\dim I + (p-1)n$, where
 $p\in\set N$ is a quantity they call the ``polynomial growth'' of the
 ideal~$I$. Not much is known about this quantity. It seems that we have $p=1$
 in most cases of practical interest, but we do not know whether it is connected
 to more classic quantities defined for (operator) ideals, or even how to
 compute it for a given ideal~$I$.

 \begin{problem}
   Clarify the meaning of ``polynomial growth'' introduced in~\cite{chyzak09a}.
   Can the definition in~\cite{chyzak09a} be replaced by another one, possibly not equivalent, which also
   satisfies the bound on $\dim I$ stated above?
   Is there an efficient algorithm for computing the polynomial growth of a given operator ideal?
 \end{problem}

 For sums involving Stirling numbers, it would also be conceivable to have a
 creative telescoping algorithm that exploits the special form of their
 generating function. For example, for the Stirling numbers of the second kind,
 $\sum_{n,k=0}^\infty S(n,k)\frac{x^n}{n!} y^k=\exp(y(\e^x-1))$ is not D-finite
 but still elementary, so generalized techniques as discussed in
 Section~\ref{sec:2} might apply. The function $f(x,y)=\exp(y(\e^x-1))$ is also
 an example of a function satisfying a system of algebraic differential equations (ADE):
 we have $f_x(x,y)-y f_y(x,y)-yf(x,y)=0$ and $f(x,y)f_{y,y}(x,y) - f_y(x,y)^2 = 0$.
 Other prominent examples of non-D-finite functions satisfying algebraic differential
 equations are the generating function for the partition numbers $\prod_{n=0}^\infty(1-x^n)^{-1}$
 and the Weierstra\ss\ $\wp$-function. Solutions of ADEs also appear in combinatorics, for
 example as generating functions of certain restricted lattice walks~\cite{bernardi16}.

 While there is a reasonably well developed elimination theory for systems of algebraic
 differential equations~\cite{mansfield93,gerdt97,hubert00,chen03, gao2013}, no creative telescoping algorithm for this class
 of functions is known.

 \begin{problem}
   Develop a creative telescoping algorithm applicable to functions satisfying
   systems of ADEs.
 \end{problem}

 For approaching this problem, it may become appropriate to adapt the
 specification of the creative telescoping problem. In a context where
 quantities are defined by non-linear equations, it may be too restrictive to
 require that the telescoper be a linear operator. On the other hand, allowing
 non-linear operators as telescoper does not seem sensible either as long as the
 main motivation for creative telescoping is to derive equations for definite
 integrals, because the application of an integral operator does in general not
 commute with such an operator. It is a part of the
 problem to determine a suitable adaption of the creative telescoping problem.

 \section{The Inverse Problem}

 Using creative telescoping, we can obtain a recurrence satisfied by a given
 definite sum. The recurrence then serves as a basis for obtaining further
 information about the sum, such as asymptotic estimates or closed from
 expressions. The classical application is to use Zeilberger's algorithm in
 combination with Petkovsek's algorithm~\cite{petkovsek92,petkovsek97} in order
 to decide whether a given definite hypergeometric sum admits a hypergeometric
 term as a closed form. If the sum comes from some application, there is a
 certain chance that such a representation exist. However, an arbitrarily chosen
 sum is not likely to have a closed form.  It is even less likely for an
 arbitrary recurrence (which may or may not come from creative telescoping) to
 have a hypergeometric closed form. People have therefore designed algorithms
 for finding more general types of closed form solutions of recurrence
 equations, for example d'Alembertian solutions~\cite{abramov94,petkovsek97} or
 liouvillean solutions~\cite{put97,hendriks98}. Even more generally, we could
 ask whether a given recurrence admits a solution that can be expressed as a
 definite sum. In a way, this would be the inverse problem of creative
 telescoping. Chen and Singer in~\cite{chen12d} gave a characterization of possible linear operator that can be
 minimal telescopers for bivariate rational functions. However, no algorithm is known for solving this problem in the general case,
 but it would be very valuable for practical applications.

 \begin{problem}
   Design an algorithm which takes as input a nonzero recurrence operator $P\in\set Q[n][S_n]$
   and finds, if at all possible, a bivariate hypergeometric term $f(n,k)$ which has $P$
   as a telescoper.
 \end{problem}

 The analogous problems for the differential case and the two mixed cases are interesting as well.

 In recent years there has been some activity by van Hoeij and
 collaborators concerning solutions of recurrences or differential equations in
 terms of hypergeometric
 series~\cite{hoeij10,cha10,cha10a,kunwar13,imamoglu15}. In a way, these
 algorithms solve only special cases of the inverse problem for creative
 telescoping, thus indicating perhaps that the general problem may be very
 difficult.

 \section{Computational Challenges}

 Creative telescoping is not only of theoretical interest but it is also a valuable tool in all contexts
 where summation and integration problems arise that are beyond the scope of any reasonable hand-calculation.
 For example, the proof of the qTSPP conjecture~\cite{koutschan10a}, which was obtained using Koutschan's Mathematica
 package~\cite{koutschan10c},
 involves a creative telescoping problem that leads to a certificate of 4Gb size. Such computations are clearly
 not feasible by hand, and they are also challenging for computers. We shall therefore conclude this paper
 with two explicit computational challenges which to our knowledge are not feasible by any software currently
 available.

 The first problem is quoted from~\cite{kauers11b} and concerns the computation of diagonals. Again, the diagonal
 of a multivariate series
 $\sum_{n_1,\dots,n_d=0}^\infty a_{n_1,n_2,\dots,n_d}x_1^{n_1}\cdots x_d^{n_d}$ is defined as
 $\sum_{n=0}^\infty a_{n,n,\dots,n} x^n$. The diagonal of a D-finite series is D-finite~\cite{lipshitz88}, and creative
 telescoping can be used, at least in principle, to derive a recurrence for the diagonal terms $a_{n,n,\dots,n}$
 from a given set of defining equations for the original multivariate series.
 \begin{problem}
   For $d=4,\dots,12$, prove recurrence equations for the diagonals of the
   rational series $1\Big/\Bigl(1 - \sum_{i=1}^d\frac{x_i}{1-x_i}\Bigr)$
   conjectured in~\cite{kauers11b}.
 \end{problem}
 For $d=1,2$, the problem is easy. For $d=3$, it was solved in~\cite{bostan11}.

 In 2002, Beck and Prixton made an effort to compute the Ehrhart polynomial of Birkhoff polytopes~\cite{beck03}, a
 quantity that is relevant in discrete geometry~\cite{beck07}. There is a Birkhoff polynomial associated to every
 $n\in\set N$. They succeeded in computing the full Ehrhart polynomial for all $n\geq9$, and the most significant
 coefficient for the case $n=10$. As a computational challenge, we pose the computation of the
 full Ehrhart polynomial for $n=10,11,12$. We take advantage of Theorem~2 of~\cite{beck03}, where these polynomials
 are expressed as integrals that are amenable to creative telescoping.
 \begin{problem}
   For $n=10,11,12$, compute the polynomial
   \[
   H_n(t)=\frac1{(2\pi i)^n}\int_{|z_1|=\epsilon_1}\cdots\int_{|z_n|=\epsilon_n} (z_1\cdots z_n)^{-t-1}
      \biggl(\sum_{k=1}^n\frac{z_k^{t+n-1}}{\prod_{j\neq k}(z_k-z_j)}\biggr)^n dz_n\cdots dz_1,
   \]
   where $0<\epsilon_1,\dots,\epsilon_n<1$ are arbitrary.
 \end{problem}
 This problem is similar to the previous one in so far as diagonals can be rephrased as contour integrals. But it
 is different in that we ask for the polynomials $H_n(t)$ rather than for some differential equation
 satisfied by~$H_n(t)$. Following the standard approach, we would first use creative telescoping to compute such
 a differential equation, then determine the space of polynomial solutions of this equation, and then find
 the unique element of this space that matches the initial terms of~$H_n(t)$. This element must be $H_n(t)$ itself.
 In the present context, this approach may not be feasible because the computation of the first coefficients of $H_n(t)$
 is not much easier than the computation of the whole polynomial. So part of the question is whether creative telescoping
 can help to compute the polynomials directly, without the detour through a differential equation.

 \bibliographystyle{plain}
 \bibliography{bib}

\end{document}